\documentclass[11pt]{article}
\usepackage{graphicx}

\def\be{\begin{equation}}
\def\ee{\end{equation}}
\def\ba{\begin{eqnarray}}
\def\ea{\end{eqnarray}}

\def\12{{1\over 2}}

\def\msun{M_\odot}

\def\etal{{\it et~al.~}}
\def\ltsima{$\; \buildrel < \over \sim \;$}
\def\simlt{\lower.5ex\hbox{\ltsima}}
\def\gtsima{$\; \buildrel > \over \sim \;$}
\def\simgt{\lower.5ex\hbox{\gtsima}}

\begin{document}

\title{\bf Evolution of the First Supernovae in Protogalaxies: \\
Dynamics of Mixing of Heavy Elements\footnote{This paper is published
in Astronomy Reports, 2012, Vol. 56, No. 12, pp. 895.}}
\author{E.~O.~Vasiliev$^{1,2}$\thanks{eugstar@mail.ru}, E. I. Vorobyov$^{3}$, E.~E.~Matvienko,$^2$ A.~O.~Razoumov$^4$,
Yu.~A.~Shchekinov$^{2}$ \\
\it $^1$Institute of Physics, Southern Federal University, Rostov-on-Don, Russia \\
\it $^2$Physics Department, Southern Federal University, Rostov-on-Don, Russia \\
\it $^3$Institute of Astronomy, University of Vienna, Austria \\
\it $^4$SHARCNET/UOIT Consortium, Oshawa, Canada}

\date{}

\maketitle

\begin{abstract}
The paper considers the evolution of the supernova envelopes produced by Population III
stars with masses of $M_*\sim 25-200~M_\odot$ located in non-rotating protogalaxies with masses of $M\sim 10^7~M_\odot$
at redshifts $z=12$, with dark-matter density profiles in the form of modified isothermal spheres. The
supernova explosion occurs in the ionization zone formed by a single parent star. The properties of
the distribution of heavy elements (metals) produced by the parent star are investigated, as well as the
efficiency with which they are mixed with the primordial gas in the supernova envelope. In supernovae with
high energies ($E\simgt 5\times 10^{52}$~erg), an appreciable fraction of the gas can be ejected from the protogalaxy,
but nearly all the heavy elements remain in the protogalaxy. In explosions with lower energies ($E\simlt
3\times 10^{52}$~erg), essentially no gas and heavy elements are lost from the protogalaxy: during the first one to
threemillion years, the gas and heavy elements are actively carried from the central region of the protogalaxy
($r\sim 0.1~r_{vir}$, where $r_{vir}$ is the virial radius of the protogalaxy), but an appreciable fraction of the mass of metals subsequently returns when the hot cavity
cools and the envelope collapses. Supernovae with high energies ($E\simgt 5\times 10^{52}$~erg) are characterized by
a very low efficiency of mixing of metals; their heavy elements are located in the small volume occupied by
the disrupted envelope (in a volume comparable with that of the entire envelope), with most of the metals
remaining inside the hot, rarified cavity of the envelope. At the same time, the efficiency of mixing of heavy
elements in less energetic supernovae ($E\simlt 3\times 10^{52}$~erg) is appreciably higher. This comes about due to
the disruption of the hot cavity during the collapse of the supernova envelope. However, even in this case,
a clear spatial separation of regions enriched and not enriched in metals is visible. During the collapse of
the supernova envelope, the metallicity of the gas is appreciably higher in the central region ($[Z]\sim -1$ to 0) than at the periphery ($[Z]\sim -2$ to $-4$) of the protogalaxy; most of the enriched gas has metallicities
$[Z]\sim -3.5$ to $-2.5$. The masses of enriched fragments of the supernova envelope remain appreciably lower
than the Jeans mass, except in regions at the center of the protogalaxy upon which the surrounding enriched
gas is efficiently accreted. Consequently, the birth of stars with metallicities close to those characteristic of
present-day Galactic stars is very probable in the central region of the protogalaxy.
\end{abstract}



\section{Introduction}

\noindent

The first stars in the Universe began to be born
in protogalaxies with total masses $M\sim 10^7~M_\odot$ at
redshifts $z \sim 12$ [1, 2]. The initial mass function of
these first stars was probably shifted toward more
massive stars due to the low efficiency of cooling
when $T\simlt 100$~K and the low opacity of the primordial
gas [3]. The masses of the first stars probably lay in
the interval from tens to hundreds of solar masses [4],
making their lifetimes very short -- in all, several million
years [5]. After the numerical computations of [6],
it has often been suggested that only one star initially
arose in low-mass protogalaxies, although a group of
stars could form under some conditions [7-10].

The first stars exerted an appreciable influence on
the ambient medium. First, ionizing photons emitted
by the first stars formed ionization zones, substantially
redistributing the gas in their vicinity [11-18].
Second, the first stars exploded as supernovae (SN)
at the end of their lives, ejecting heavy elements (metals)
produced in their centers during their lifetimes
into the surrounding space [19-26]. The number of
ionizing photons, SN energy, and mass of ejected
heavy elements depend on the mass of the star [27,
29]. The energy of SN explosions of massive stars
could be sufficient to disrupt a protogalaxy [30, 31]
and eject a substantial quantity of metals into intergalactic
space [20]. The explosion of multiple SN will
facilitate the formation of such outflows [32, 33]. Gas
that has been enriched in heavy elements could be
accreted onto neighboring protogalaxies and, when
a metallicity above a certain critical value has been
obtained [34-37], stimulate the formation of stars
with a mass function close to the current one [23, 38].
Since stars can be born in a medium with any metallicity,
transition-population stars with extremely low
metal contents could be born in enriched gas [39, 40];
such stars could carry information about the chemical
composition of the products of nucleosynthesis in the
first stars [41, 42]. Low-mass stars with extremely
low iron abundances  ${\rm [Fe/H]}\sim -5$ have recently been
discovered in the Galaxy [43-46], some of which
have appreciably higher abundances of other elements,
e.g., ${\rm[O/Fe]}\sim 2$ [47, 48]. However, stars with
extremely low abundances of all elements ${\rm [X/H]}\sim -5$ 
to $-4$, are also encountered [49]. These latter
may be considered transition-population stars, but
the question of their origin remains open [42].

Generally speaking, the spatial distribution of
heavy elements (mixing) depends substantially on
the relative flows of gas containing metals. When
numerous shocks are present, the regions containing
enhanced metal contents will be disrupted, and the
distribution of heavy elemetns becomes more uniform
(total mixing) [50]. If these relative motions die out in
themedium, themetals will end up being contained in
compact “islands”, which are not subsequently disrupted
(incomplete mixing) [51, 52]. It is obvious that
the effect of incomplete mixing will be enhanced in
a radiatively cooled medium [51]. The characteristic
time for the expansion and cooling of the envelopes of
the first supernovae is fairly short, comprising several
million years. There are no other sources to support
the relative motions of gas during the explosion of
a single star, and the gas motions die down as a
consequence of efficient cooling in the envelope. As a
result, heavy elements that are initially concentrated
in the SN ejecta are poorly mixed with the entrained
gas of the envelope [53]. Incomplete mixing could
lead to the possible birth of stars with metallicities
close to, or even exceeding, modern values. Moreover,
since the energy of the explosion and the mass of
ejected heavy elements depends on the mass of the
star, the efficiency of mixing in early galaxy will be
different following explosions of stars with different
masses. Of course, when a star cluster exists at the
center of a protogalaxy, a series of SN explosions
is possible, whose energy could support the relative
gas motions for a more prolonged time and facilitate
more efficient mixing of the metals. However, the low
efficiency of fragmentation in the primordial gas is not
favorable for the birth of groups of stars.

Other factors that could lead to an enhancement
of mixing could include mergers with lower-mass
protogalaxies and the accretion of intergalactic gas,
leading to turbularization of the gas in protogalaxies.
Indeed, in a heirarchical model for the formation of
structure, large objects appear as a result of mergers
of smaller objects [54-57]; therefore, mixing can
occur as the mass of the galaxy grows, when mergers
or tidal and accretion flows encompss a mass comparable
to the total mass of the galaxy. At later stages in
the growth of a galaxy mass due to mergers, it must
be taken into account that mergers with low-mas
objects will exert only a local effect. However, on long
time scales, the regions of mixing will cover virtually
the entire galactic disk, under the action of mergers.
Mergers with protogalaxies with comparable masses
occur rarely.the characteristic time between such
merger events turns out to be appreciably longer than
the time scale for the expansion and cooling of the SN
envelope. The time for the accretion of intergalactic
gas after virialization of the protogalaxy and the
formation of the first stars is probably also longer
than the cooling time for the SN envelope, due to the
depletion of accreting material after virialization [58].
We do not consider the role of mergers and accretion
in mixing of metals in more detail here. In the
current paper, we restrict our analysis to the evolution
of envelopes formed by single supernovae and
study the efficiency of mixing of metals. Our computations
assumed a $\Lambda$CDM cosmological model:
$(\Omega_0,\Omega_{\Lambda},\Omega_m,\Omega_b,h ) = (1.0,\ 0.76,\ 0.24,\ 0.041,\ 0.73 )$;
the relative concentation of deuterium was taken to
be $n[{\rm D}]/n = 2.78\times 10^{-5}$ [59].


\section{A model of a protogalaxy and numerical methods}

\noindent

In this section, we provide a brief description of
themain parameters of the model, numerical methods
(described in more detail in [10, 18, 60]), and initial
conditions.

\subsection{Main Parameters}

In the model considered, a protogalaxy consists of
gas surrounded by a spherically symmetrical halo of
dark matter. The dark-matter density profile has the
form of a modified isothermal sphere:
\be
 \rho_h(r) = {\rho_0 \over 1 + (r/r_0)^2},
 \label{halo}
\ee
where $r_0$ and $\rho_0$ are the radius of the core and the
central density, and the total (dark halo and gas) mass
of the protogalaxy $M_h$ is taken to be $10^7~M_\odot$.
For redshift $z = 12$, this corresponds to $3\sigma$
perturbations in a $\Lambda$CDM model having parameter values
equal to those derived from the results of three years
of observations of the cosmic microwave background
(CMB) radiation with the WMAP satellite [59]. The
virial radius of such a protogalaxy is $r_v = 520$~pc (see
the virial relations, for example, in [61]). It follows
from simple estimates [1, 2] that a protogalaxy with a
mass of $10^7~M_\odot$ at redshift $z = 12$ is efficiently cooled,
giving rise to the conditions required for the birth of
the first generation of stars.

Since the minimum temperature of the primordial
gas at high redshifts varies from 40 to 200~K, which
facilitates the efficient formation of H$_2$ and HD
molecules [2], the rate of accretion onto the protostellar
core is higher than for the modern chemical
composition of the gas ($\dot M \sim c_s^3/G$ [62]). Therefore,
first-generation stars were substantially more
massive than stars of subsequent generations. It
follows from numerical models that their masses
could vary within wide limits, $\sim 10-10^3~M_\odot$ (see, e.g.,
[61]). We studied the evolution of the ionization zones
surrounding stars with masses of 25, 40, 120, and
200~$\msun$ in [18]. This choice of masses was dictated
by the fact that such stars explode as supernovae,
ejecting the products of stellar nucleosynthesis into
the surrounding gas, rather than collapsing directly
into black holes [27].

The explosions of the first SN enriched the surrounding
primordial gas in heavy elements (metals).
The energy of the explosion and mass of metals produced
during the lifetime of the star depends on the
starЎЇs initial mass (Table). The presence of metals
in the gas and their relative concentration determines
the efficiency of star formation and the parameters of
the stars that are born. Therefore, the redistribution
of metals after the SN explosion is of fundamental
importance for our understanding of the entire subsequent
history of the evolution of the galaxy and its
stellar populations. We present here computations
for several models, whose parameters are presented
in the Table. Note that the evolution of a star with a
mass of 25~$\msun$ could proceed along at least two paths,
and eject $\sim 2.1\msun$ or $\sim 3.3\msun$ of metals (models 25
and 25F in [63], respectively). The explosion energy
for massive stars (140-260$\msun$) is often assumed to
be the same and equal to $10^{53}$~erg, although it was
shown in [27] that this energy grows with mass from
$\sim 10^{52}$ to $10^{53}$ erg. Since there are no data on the
number of photons emitted by a star with a mass of
140~$\msun$ in [5], we used the data for a star with a mass
of 120~$\msun$ to obtain the distribution of the gas before
a SN explosion of a star with this mass.

\begin{table}[!ht]
\caption{Main characteristics of the first supernovae}
\center
\begin{tabular}{ccccc}
\hline
\hline
   Stellar mass, $M_\odot$  & Ejecta, $M_\odot$  & Mass of metals, $M_\odot$  & Explosion energy, erg & Refs  \\
\hline
25   &   $\sim 21$       & $\sim 2.1$        &  $10^{51}$         & \cite{ww95}   \\
25   &   $\sim 22$       & $\sim 3.3$        &  $10^{52}$         & -- \\
40   &   $\sim 34$       & $\sim 8.2$        &  $3\times 10^{52}$ & --   \\
140  &   $\sim 140$      & $\sim 63$         &  $10^{52}$         & \cite{heger02} \\
200  &   $\sim 200$      & $\sim 98.5$       &  $5\times 10^{52}$ & -- \\
\hline
\hline
\end{tabular}%
\label{table1}
\end{table}

\subsection{Numerical Scheme}

The gas-dynamical equations were solved numerically
in cylindrical coordinates under the approximation
of axial symmetry ($z,r,\phi$) using a finitedifference
method with separation of operators [64].
A piecewise, parabolic, third-order interpolation
scheme was used to transport the gas [65].

To obtain the initial distribution of the gas density
in the external gravitational potential $\Phi_{h}$ of the dark
matter, the equilibrium equation was solved numerically
in cylindrical coordinates, $(z, r)$. The gas was
assumed to be neutral, to have a molecular weight
$\mu = 1.22$, and to be isothermal, with a temperature
equal to the virial temperature $T_{\rm vir}(M)$. Themethod
used to solve the equilibrium equations is described
in [66]. The interations were continued until the gasdensity
profile corresponded to the mass of the gas
inside the virial radius, $M_{\rm g}= (\Omega_b / \Omega_m ) M_{\rm h}$. In our
current study, we restrict our consideration to the
evolution of spherical (non-rotating) protogalaxies.

It is assumed in our model that all chemical reactants
(atoms, ions, molecules) move like passive
components; i.e., they have the same velocity field as
the gas, making it possible to restrict our solution
to the equations of motion of the gas. The chemical
kinetics of the primordial gas include the following
main components: H, H$^+$, H$^-$, He, He$^+$, He$^{++}$, H$_2$, H$_2^+$, D, D$^+$ and HD.
To compute the electron
concentration, we assumed conservation of charge.
The mass fraction of helium was taken to be $Y_{\rm He} = 0.24$.

We considered two temperature intervals in our
chemical-kinetics computations: high-tem\-pe\-ra\-tu\-re
($T>2\times 10^4$~K) and low-temperature ($T<2\times 10^4$~K). 
At the high temperatures, the concentrations
of ions and molecules were calculated from their
equilibrium values, while a system of equations for
the non-equilibrium chemical kinetics was solved in
the low-temperature case. The rates of chemical
reactions were taken from [67-69]. The chemical kinetics
equations were solved using a fifth-order
Runge–Kutta–Cusp method with automatic step
selection [70].

The energy equation took into account radiative
losses characteristic of the primordial gas: cooling
during recombination, collisional excitation of hydrogen
and helium, free–free transitions, Compton
interactions with CMB photons [71], and molecular
cooling of H$_2$ [69] and HD [72, 73]. The contribution
of metals should also be included for gas with
non-zero metallicity. The cooling function in the
high-temperature region was calculated for a collisional
gas with a metallicity of $(0.001-4)~Z_\odot$ using
the method of [74] and is presented as a table of
$\Lambda(T,Z)$ values (about 20 metallicities and about 100
temperatures). In the low-temperature region, it
was assumed that metals contributed to cooling via
energy losses in fine-structure lines of carbon and
oxygen [36]. The kinetics of these elements was not
calculated separately, and it was assumed that all the
carbon was singly ionized and that the oxygen was
neutral, which is fully acceptable for the conditions
considered [36]. Moreover, we included the effect of
interactions between molecules, ions, and atoms of
metals with CMB photons [75-77]: when gas temperatures
close to the CMB temperature are reached,
molecules, ions, and atoms of metals can be excited
by CMB photons, and may transfer energy to the gas
during collisions, heating this gas.

\subsection{Initial Conditions}

In our computations, we assumed that the dark
matter in a protogalaxy at redshift $z = 12$ is already
virialized, so that it has achieved a configuration described
by (1). The protogalaxy gas cools, and the gas
concentration in the central region grows, reaching
$\sim 10^8$~cm$^{-3}$ in our computations. Further, we assume
that a star is born in the center of the protogalaxy.

The ionizing photons from the star form a surrounding
ionization zone, which encompasses an appreciable
volume in the protogalaxy by the end of the
star’s lifetime [18]. As initial conditions before the
SN explosion, we took the distributions of the density,
temperature, and chemical components obtained in
our computations [18] at the end of the lifetime of a
star with the corresponding mass. A thermal energy
corresponding to the explosion energy and a mass of
metals are added to a central region with a radius of
two parsec (Table).

We used a non-uniform computational grid to
improve the resolution at the center of the protogalaxy
[10]. The density of the grid can be controlled
using a coefficient $A$: decreasing $A$ leads to
an increase in resolution in the inner regions of the
computational grid and a decrease in resolution in
outer regions. In our computations, we adopted $A = 1.5$ 
and used a $900\times900$ grid, which corresponds to a
physical resolution better than 0.1~pc in a region with
a radius of 10~pc, which degrades to $\sim 1$~pc at a radius
of $r\simeq200$~pc. The inhomogeneous division of the grid
was the same in the $z$ and $r$ directions.


\section{Evolution of gas and metallicity}

\noindent

In a high-energy SN explosion ($E\sim10^{53}$~erg) in
a protogalaxy with mass $M\simlt 10^7\msun$, the kinetic
energy of the expanding gas envelope is sufficient
for it to reach the virial radius of the protogalaxy, so
that the protogalaxy is essentially disrupted [30, 31].
The gas in such objects will be essentially fully swept
out by the SN explosion, and the protogalaxy will
be filled with the hot ($T\sim 10^5-10^6$~K), rarified 
($n\sim 10^{-4}-10^{-3}~$cm$^{-3}$), very high-metallicity gas ejected
by the SN, which occupies a volume corresponding
to more than half the virial radius of a protogalaxy
with mass $M\sim 10^7~M_\odot$. The metallicity of
this gas is determined by the mass of the star, and
can exceed the solar value by more than an order
of magnitude -- practically half the mass of the SN
can be reprocessed into heavy elements [27]. The
cooling time in the hot gas, which is of the order of
several million years, turns out to be comparable to
or longer than the time scale for the loss of an appreciable
mass of gas (about 50\%), since the remaining
gas is located at the periphery of the galaxy (at distances
exceeding the virial radius), and its velocities
are comparable to the escape speed [60] (see also
our figure for the SN model with $M = 200~M_\odot$ and
$E = 5\times 10^{52}$~erg). Such “dark” (with a low mass
of gas) dark-matter halos will merge with lowermass,
gas-rich protogalaxies in which star formation
has not occurred, and in which a sufficient mass of
gas for star formation may accumulate in the future.
Note also that such “dark” halos could be “invisible”
satellites of galaxies: there appears to be a
discrepancy between the numbers of theoretical and
observed satellites in the Galaxy —- the so-called
“missing-satellites problem” [78]. Another consequence
of a high-energy SN explosion in a low-mass
halo could be modification of the dark-matter profile
in the protogalaxy, such as a transition from a cusp to
a flat profile [79-81].

The masses of the first stars could vary within fairly
broad limits, from several tens to several hundreds of
solar masses (see, e.g., [61]). Of course, the energy
of the SN depends on the mass of the star and varies
in the range $E\sim10^{51}-10^{53}$~erg [27]. Consequently, the
kinetic energy of the gas in the SN remnant may be
insufficient to overcome the gravitation of the dark
halo of the protogalaxy, so that all (or a large fraction
of) the gas remains inside the protogalaxy. The hot
cavity\footnote{
We will understand the cavity to be the region containing
the hot gas in the central region of the protogalaxy, which
contains part of the unmixed ejecta). The ejecta is the 
heavy-element-enriched gas ejected from the parent star. 
Subsequently, the accumulated primordial gas in the SN envelope
mixes with the highly enriched gas, and so will contain part
of the ejecta.} gradually cools, the pressure is lowered, and
part of the gas in the SN envelope begins to move
toward the center of the protogalaxy due to the gravitational
field of the dark halo; i.e., the collapse of the
SN envelope begins. It is natural to expect that the
primordial, or metal-poor, gas in the envelope mixes
with the metal-rich gas of the ejecta.

Figures 1 and 2 show the distributions of the
density and temperature of the gas, the relative concentration
of molecular hydrogen, and the metallicity
at times t = 1, 5, 10, 15, 20~Myr (from top to
bottom) after the explosion of a SN with an initial
mass of 140~$M_\odot$ and an energy of $10^{52}$~erg in the
center of a spherical protogalaxy with a total mass
of $M = 10^7~\msun$. The SN envelope has formed by
time $t = 1$~Myr, when it almost reaches a radius of
100~pc, which corresponds to about one-fifth of the
virial radius. Small perturbations of the density in the
envelope due to the Rayleigh-Taylor instability can
be seen. These perturbations lead to fragmentation
and the almost total disruption of the envelope, as is
visible in the lower maps in Fig. 1. Since the gas in
the envelope cools efficiently, thermal instability also
plays a role in the fragmentation of the gas. Extended
“tongues” with lower density and higher temperature
had already formed during the existence of the parent
star, due to penetration of the ionization zone [18].
Due to the enhanced degree of ionization, molecular
hydrogen forms efficiently in the SN envelope and
remnant HII zone, which cools the gas when $T\simlt 10^4$~К. 
However, this is not the only source of cooling
of the gas in the SN envelope. A small fraction of
the heavy elements of the ejecta gradually become
mixed with the primordial, swept up gas at the inner
side of the envelope. We can clearly see that the
efficiency of the mixing is low, since all the heavy
elements are contained in the hot cavity (upper maps
in Fig. 1). At later times, heavy elements can be seen
in disrupted fragments of the envelope (subsequent
maps of Fig. 1).

\begin{figure}
\center 
\includegraphics[width=12cm]{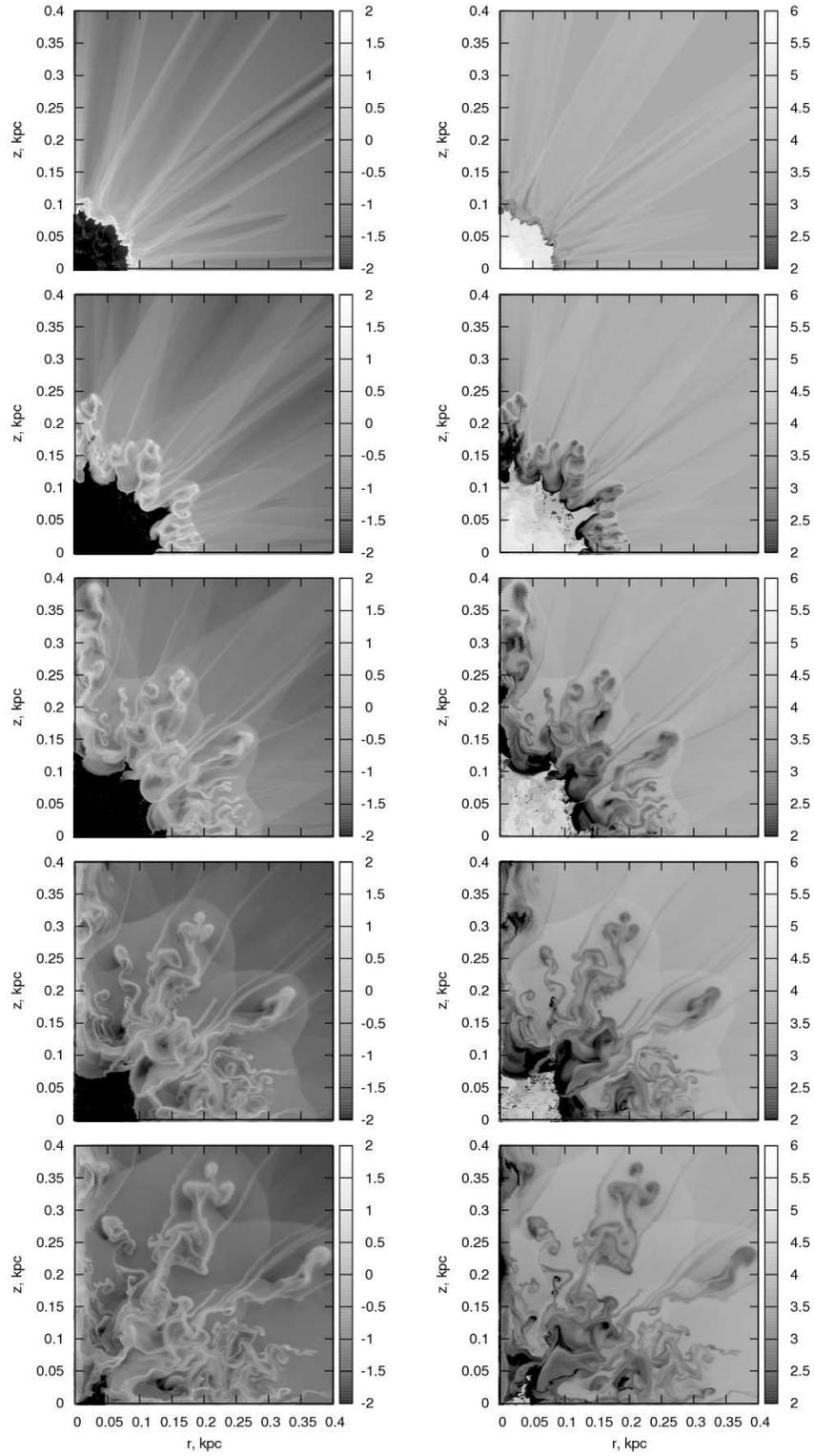}
\caption{
Distribution of the density (left) and temperature (right) of the gas at times $t = 1, 5, 10, 15, 20$~Myr after
the explosion of a star with mass 140~$M_\odot$ and energy $10^{52}$~erg (from top to bottom) in the central region of a protogalaxy with a total mass $M = 10^7~\msun$.
}
\label{mapnt}
\end{figure} 

\begin{figure}
\center 
\includegraphics[width=12cm]{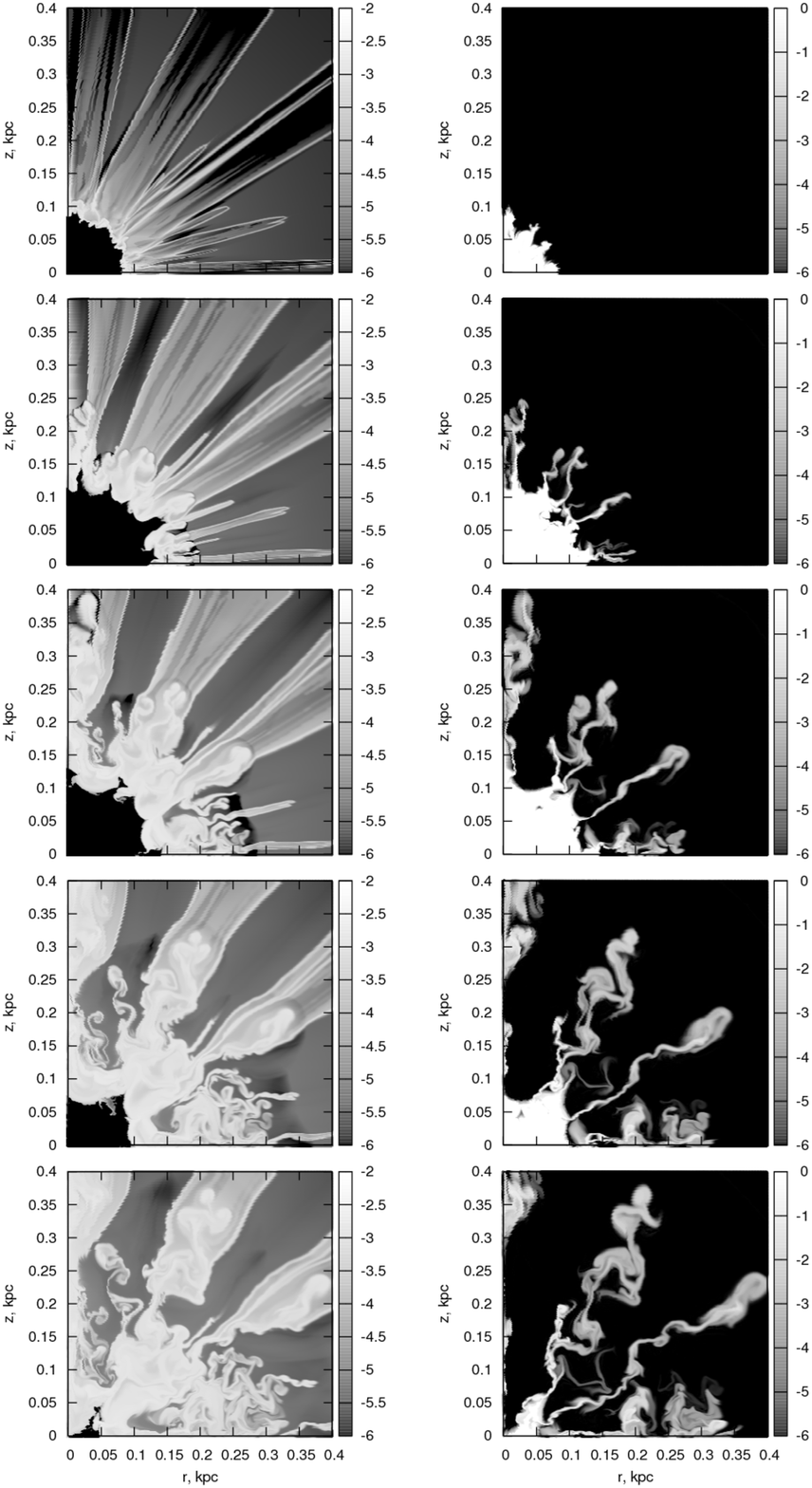}
\caption{
Same as Fig. 1 for the relative concentration of molecular hydrogen (left) and the metallicity (right) of the gas.
}
\label{maph2m}
\end{figure} 

As was indicated above, with time, the pressure of
the hot gas in the cavity is reduced by cooling, and
the SN envelope begins to collapse under the action
of the gravitation of the dark halo; this is visible in
Fig. 1 for $t = 5$~Myr (third row of maps from the top).
This is more clearly visible for the subsequent time
$t = 10$~Myr. Some envelope fragments continue to
move away from the center, but part of the envelope
moves toward the center, gradually mixing with the
hot, metal-rich gas in the cavity. We can clearly
see that H$_2$ (Fig. 2), and then HD, molecules form
efficiently in the envelope fragments, the gas cools
rapidly, and the temperature falls below 100~K in
some regions (Fig. 1). The gas flows converging on
the central region of the protogalaxy form a series
of shocks and near-acoustic waves, which facilitate
mixing of the hot, high-metallicity gas of the cavity
with the cool, primordial (low-metallicity) gas of the
envelope fragments. By time $t = 20$~Myr (lower maps
in Figs. 1, 2), a gravitational-compression regime is
established in the central region of the protogalaxy,
and the gas density reaches $n\sim 10^6$~cm$^{-3}$; we ceased
the computations at this point, since our model does
not include physical processes that can occur in such
a dense, metal-rich gas (the formation of CO and OH,
and the loss of energy via lines of these molecules).
It is obvious that the conditions required for star
formation will arise in this region. This possibility will
be discussed below.

Comparing the density and metallicity distributions
of the gas that have been established by a time
$t = 20$~Myr (lower maps in Figs. 1, 2), we note that
heavy elements are essentially strictly localized in
fragments of the disrupted envelope, although the
gas in some of these is still primordial. Moreover,
heavy elements are visible in “thread-like” structures,
which formed when the fragments became distinct
as a result of Rayleigh–Taylor instability. This is
a consequence of the fact that some of the highly
enriched gas of the ejecta is entrained into fragments
during their formation. The metallicities of the gas
in the fragments and these thread-like structures
are comparable, $[Z]\simeq -4$ to $-2$. An appreciable
fraction of the heavy elements is concentrated in the
central region, where the metallicity of the gas can
reach $[Z]\simeq -1$ to 0. Overall, however, the mixing
is incomplete, as is manifest by the extremely nonuniform
distribution of heavy elements: fairly enriched
regions ($[Z]\simeq -3$) and regions of primordial gas can
be located virtually right next to each other.

Overall, the evolution of the SN remnant for other
explosion energies for which there is a collapse of
the hot cavity resembles the picture described above.
Differences are associated primarily with the initial
conditions in the protogalaxy before the explosion,
i.e., the gas distribution established in the ionization
zone by the end of the star’s lifetime and the
properties of the SN -- the mass of heavy elements
ejected during the explosion and the explosion energy.
All this influences the size of the region containing
heavy elements, and therefore the efficiency of mixing
of the heavy elements. For comparison, we present
the gas density and metallicity distributions at the
times when the conditions for the formation of the
next generation of stars have arisen (a gravitational
contraction regime has been established in the central
region of the protogalaxy, the gas density reaches
$n\sim 10^6$~cm$^{-3}$).

Figure 3 shows the distributions of the gas density
(left) and metallicity (right) after the explosions of
supernovae with masses and energies $M = 25~M_\odot$
and $E=10^{51}$~erg, $M = 25~M_\odot$ and $E=10^{52}$~erg,
$M = 40~M_\odot$ and $E= 3\times10^{52}$~erg, $M = 140~M_\odot$ and
$E=10^{52}$~erg, $M = 200~M_\odot$ and $E=5\times 10^{52}$~erg
(from top to bottom) at times $t = 4.4, 13, 20, 20$,
and 16~Myr, respectively. The distributions for the
SN with high energy ($M = 200~M_\odot$ and $E= 5\times10^{52}$~erg; 
lower panels) are shown for late times, when
an appreciable fraction of the gas of the protogalaxy
has crossed the virial radius. Note that the scales
along the $y$ axes for the various distributions shown in
Fig. 3 are different. The character of the distribution
of heavy elements is the same for all the collapsed
SN envelopes: the heavy elements are poorly mixed,
and contained within strictly localized structures. For
all the models considered, the metallicities in these
structures lie in the range $[Z]\simeq-4$ to $-2$.

\begin{figure}
\center 
\includegraphics[width=12cm]{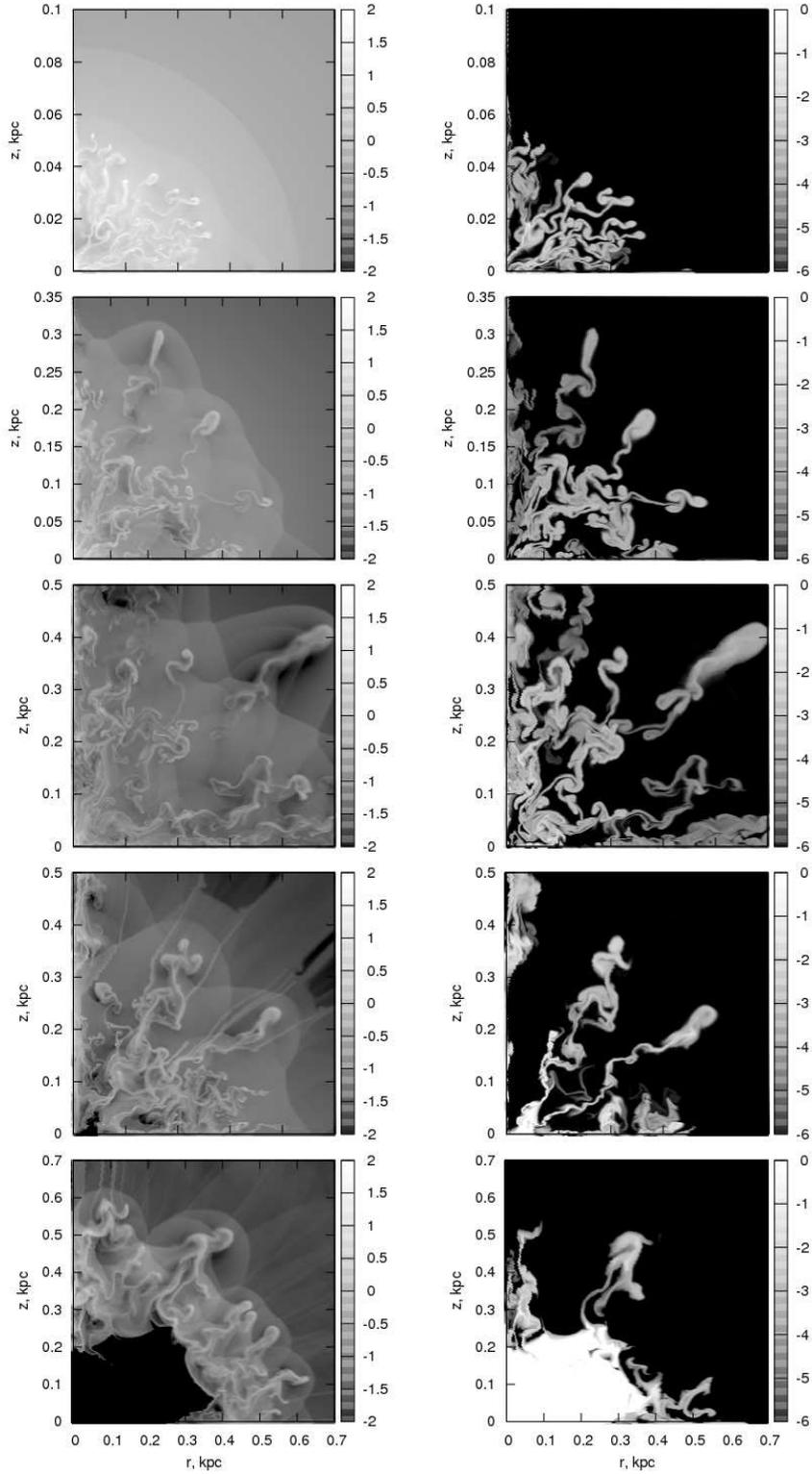}
\caption{
Distributions of the density (left) and metallicity (right) of the gas at the final times following SN explosions 
in a protogalaxy with total mass $M = 10^7~\msun$ (from top to bottom): 
$t = 4.4$~Myr for a star with mass 25~$M_\odot$ and a SN energy $10^{51}$~erg,
$t = 13$~Myr for a star with mass 25~$M_\odot$ and a SN energy $10^{52}$~erg,
$t = 20$~Myr for a star with mass 40~$M_\odot$ and a SN energy $3\times 10^{52}$~erg,
$t = 20$~Myr for a star with mass 140~$M_\odot$ and a SN energy $10^{52}$~erg,
$t = 16$~Myr for a star with mass 200~$M_\odot$ and a SN energy $5\times 10^{52}$~erg.
}
\label{mapnm}
\end{figure} 

\begin{figure}
\center 
\includegraphics[width=12.5cm]{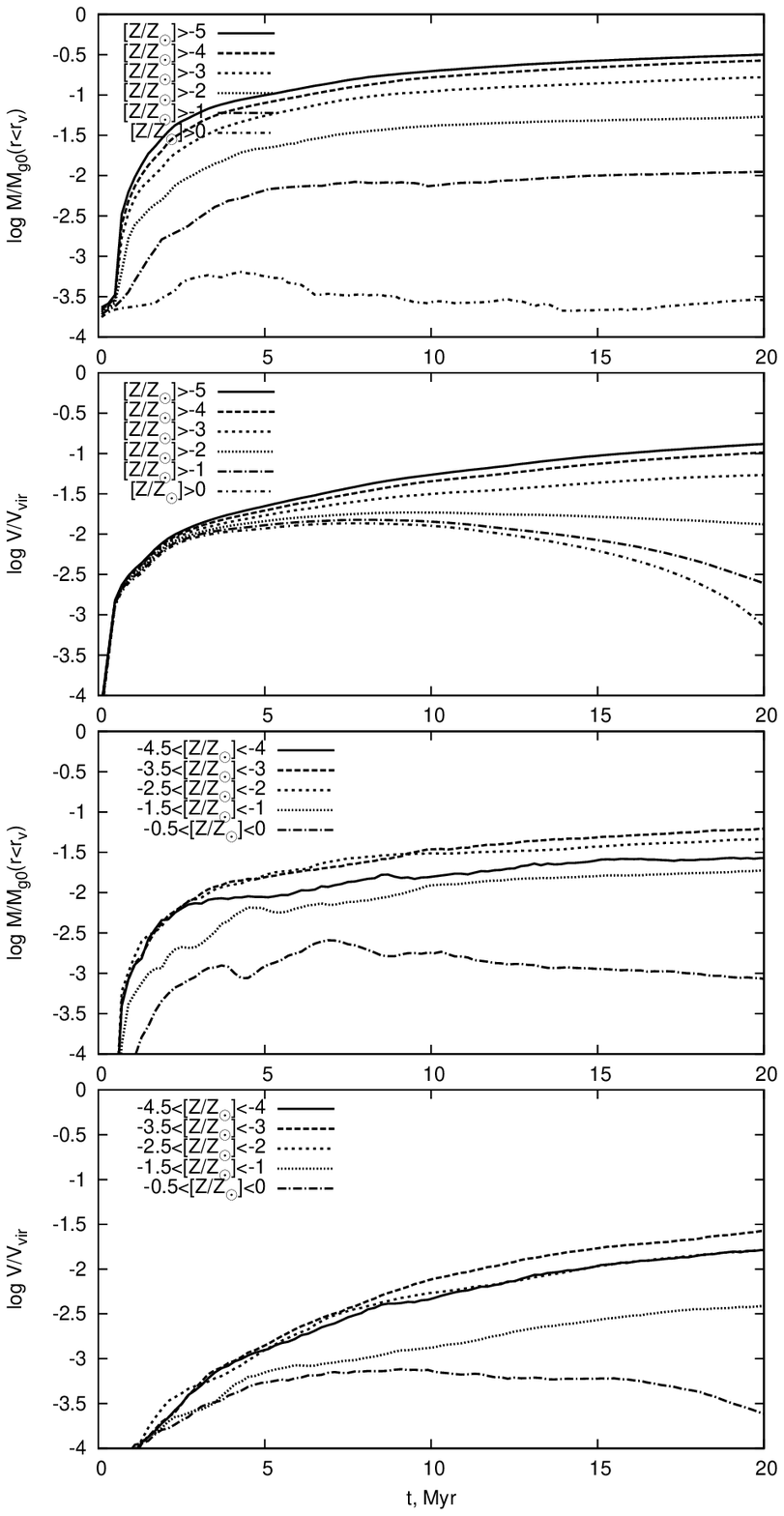}
\caption{
Evolution of the mass and volume of the gas with metallicity above a specified level (upper panels) and in intervals
(lower panels) inside the virial radius following an explosion of a star with mass 140~$M_\odot$ and an SN energy of $10^{52}$~erg.
}
\label{massz140}
\end{figure} 

The volume of gas encompassed by the shock increases.
Shortly thereafter, the volume of the enriched
gas likewise grows, as is clearly visible in Fig. 3. The
same is true for the mass of the enriched gas. As
an example, Figure 4 (two upper panels) shows the
evolution of the mass and volume of the gas with
metallicity above a specified level inside the virial
radius following a SN explosion with initial mass
140~$M_\odot$ and energy $10^{52}$~erg. The mass of enriched
gas ($[Z]>-5$) grows with time, approaching saturation
at times $t\simgt 15$~Myr. The mass and volume of gas
with metallicity $[Z]>-2$ saturates by $t\sim 7-8$~Myr.
This indicates weak mixing of this gas at subsequent
times. This gas should remain in the central region of
the protogalaxy, and the enrichment of the expanding
fragments virtually ceases. The gas in these fragments
is not enriched above a metallicity $[Z]\sim -2$.
Note that the mass of high-metallicity gas ($[Z]>0$)
remains constant with time. The volume occupied
by this gas grows over the first five million years,
then decreases. This corresponds to the initial expansion
and subsequent compression of the hot cavity.
Note here that a SN envelope with high energy 
($E\sim 5\times 10^{52} - 10^{53}$~erg) encompasses virtually the whole
of a protogalaxy with mass $10^7~\msun$, but the heavy
elements remain concentrated in the rarified, highly
enriched cavity (lower panels in Fig. 3). A substantial
fraction of the heavy elements in collapsing envelopes
is associated with dense, i.e., fairly cool (Figs. 1, 2),
gas.

Figure 4 also shows (two lower panels) the evolution
of the mass and volume of the gas in metallicity
intervals inside the virial radius following the explosion
of a SN with initial mass 140~$M_\odot$ and energy
$10^{52}$~erg. The gas with intermediate metallicities,
$-3.5<[Z]<-2$, has the greatest mass and volume,
although the number of cells with such metallicities
is not large (see the following section). This gas is
primarily concentrated in fragments of the envelope at
large distances from the center. Here, we must bear
in mind that our computations were carried out in a
cylindrical geometry; i.e., the cells at the periphery
have substantially larger volumes than those in the
central region of the protogalaxy.

As was noted above, the volume and mass of the
enriched gas depend on the SN energy, and some of
the gas can leave the protogalaxy. Figure 5 presents
the evolution of the mass of gas and heavy elements
ejected beyond radii of 0.05~$r_{v}$, 0.1~$r_{v}$, and 1.0~$r_{v}$ during
the explosion of a SN with energy $E$ and initial mass
$M$. An appreciable fraction of the gas is ejected
from the galaxy only for the SN with energy $E = 5\times 10^{52}$~erg 
and mass $M = 200~M_\odot$ (lower panels).
Roughly one-third of the mass of gas has passed
beyond the virial radius by time $t = 20$~Myr. The loss
of metals is smaller, comprising about one-tenth of
the mass of heavy elements. The remaining metals
are located primarily in the hot cavity. The cooling
time of the hot gas, which is approximately several
million years, is longer than the time scale for the
loss of an appreciable fraction of the gas mass (more
than 50\%), since the mean velocity of the gas at the
periphery of the galaxy (at radii exceeding half the
virial radius) is comparable to the escape velocity for
the protogalaxy considered. Thus, a protogalaxy with
mass $M\sim10^7~\msun$ in which high-energy supernovae
explode with with high probability become a "dark"
object with essentially no gas.

The explosion of a SN with energy $E = 3\times 10^{52}$~erg 
and initial mass $M = 40~M_\odot$ (third row
of panels from the top) is accompanied by a more
modest loss of gas (up to one-tenth of the mass of
gas). The mass of gas within radii of 0.05~$r_v$ and 0.1~$r_v$
falls sharply following the explosion, decreasing by
several orders of magnitude in the first 1.5~Myr after
the explosion. The mass of metals behaves similarly.
However, beginning from 8~Myr, the gas begins to
return to the central region of the protogalaxy -— the
SN envelope begins to collapse. By 20~Myr, the mass
of gas at distances $r < 0.05 r_v$ has actually increased
by an order of magnitude, compared to its value
before the explosion. An appreciable fraction of the
mass of metals (about 50\%) is concentrated within
$r < 0.05 r_v$ .

The loss of gas and metals from the protogalaxy
is insigificant for the remaining three models 
($M = 25~M_\odot$ and $E=10^{51}$~erg, $M = 25~M_\odot$ 
and $E=10^{52}$~erg, $M = 140~M_\odot$ and $E=10^{52}$~erg) (see the
corresponding panels in Fig. 5). The evolution of
the mass of gas and metals in the model with $M = 140~M_\odot$ 
and $E=10^{52}$~erg resembles the evolution
described for $M = 40~M_\odot$ and $E=3\times 10^{52}$~erg,
and we do not considered it in detail here. The
redistribution of gas and metals for a SN with $M = 25~M_\odot$ 
occurs within 0.05~$r_v$ and 0.1~$r_v$ for SN energies
$E=10^{51}$~erg and $E=10^{52}$~erg, respectively; up to
90\%, and for $E=10^{51}$~erg up to 99\%, of the mass of
metals is retained within the region $r < 0.1~r_v$.

Heavy elements facilitate more efficient cooling of
the gas than do H$_2$ and HD molecules [34]. The
first stars emit a substantial number of photons at
$11-13.6$~eV [5], which are capable of disrupting H$_2$
molecules if their column density is small, so that self-shielding
is not important. Therefore, the relative
number density of H$_2$ in the surrounding medium is
fairly low by the end of the star's lifetime in most of
the models considered: $\sim 10^{-6}$. The gas density in
the unstable envelope formed by the ionization front
increases and the temperature decreaes with decreasing
mass of the star and decreasing number of ionizing
photons [18]. The source of ionizing radiation
"switches off" by the time of the SN explosion, and
conditions favorable for the formation of molecular
hydrogen arise in the ionized gas of the envelope.
Therefore, the SN shock front propagates through
gas in whichmolecules are already efficiently forming.

\begin{figure}
\center 
\includegraphics[width=12.5cm]{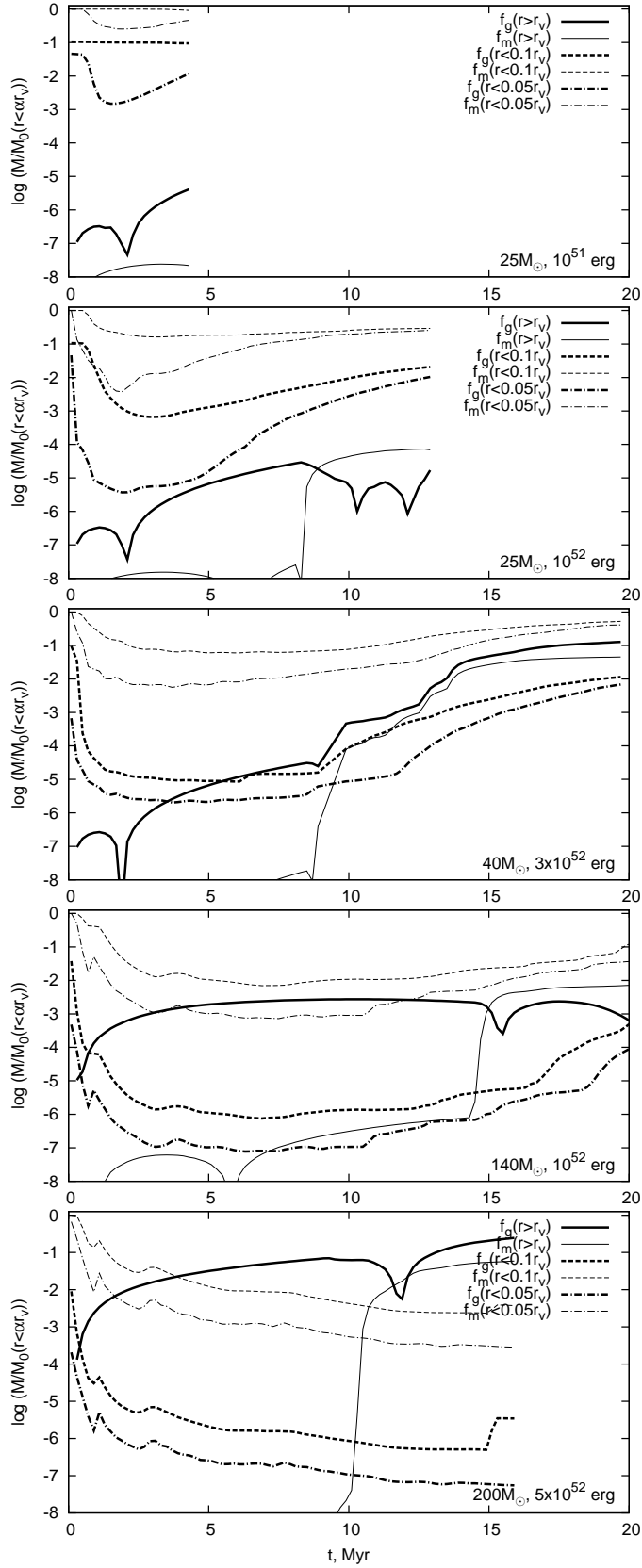}
\caption{
Mass fraction of gas $f_g$ and heavy elements $f_m$ remaining in the halo and ejected from the halo during a SN
explosion with energy $E$ and initial mass $M$. $M_0$ is the mass of gas (metals) outside or inside the radius 
$\alpha r_v$ ($\alpha = 0.05, 0.1, 1$) at the moment of the SN explosion.
}
\label{massblown}
\end{figure} 

\begin{figure}
\center 
\includegraphics[width=12.5cm]{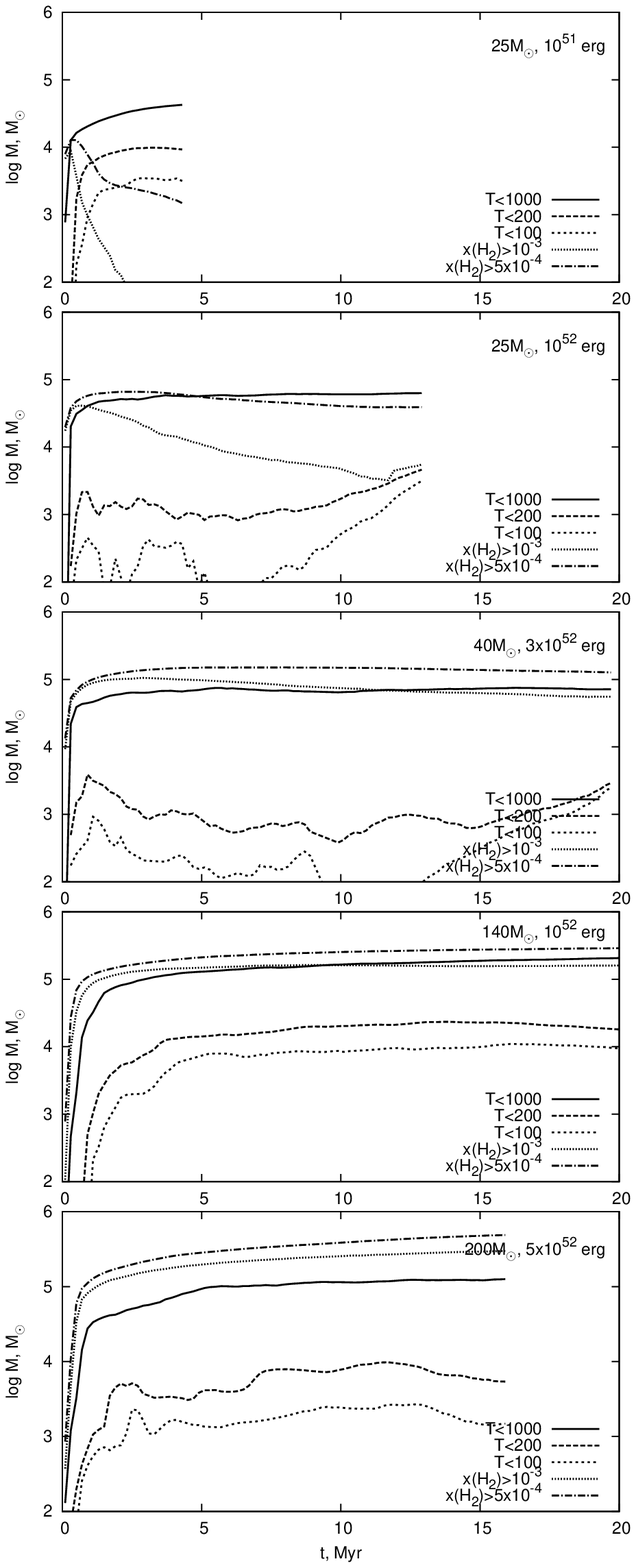}
\caption{
Evolution of the mass of gas with relative number densities of H$_2$ exceeding $5\times 10^{-4}$ and $10^{-3}$, 
and also with temperatures below 100, 200, and 1000 K, within the virial radius $r_v$ for a SN explosion with energy $E$ and initial mass $M$.
}
\label{evolt}
\end{figure} 

Figure 6 depicts the evolution of the masses of
gas with relative number densities of H$_2$ exeeding
$5\times 10^{-4}$ and $10^{-3}$, and also with temperatures below
100, 200, and 1000~K, inside the virial radius $r_v$ for
a SN explosion with energy $E$ and initial mass $M$.
Due to the high gas density in the envelope of the
ionization zone, the mass of H$_2$ grows rapidly in the
model with $M = 25~M_\odot$: the relative number density
has reached $\sim 10^{-3}$ in the densest fragments of the
envelope essentially after $0.1-0.2$~Myr. The SN shock
rapidly propagates in the hot remnant of the ionization
zone, disrupting the molecules formed there; in the
case $M = 25~M_\odot$ (two upper panels in Fig. 6), the
mass of gas with number densities higher than $10^{-3}$
falls rapidly. As the SN envelope expands, the gas in
the envelope cools, so that molecular hydrogen again
begins to form. Accordingly, the mass of gas with
molecular hydrogen number densities above $5\times 10^{-4}$
for a SN with energy $E = 10^{51}$~erg displays a modest
break at $t \sim 1.5$~Myr, and this gas mass decreases
only slightly with time in the case of a SN with higher
energy, $E = 10^{52}$~erg. In the remaining models, the
mass of molecular gas grows due to the formation of
molecules in the SN envelope and in weakly ionized
"lobes" -- remnants of the disrupted ionization-zone
envelope [18], visible in Fig. 2 as diverging "rays" with
higher molecular number densities.

In the model with $M = 25~M_\odot$ and $E = 10^{51}$~erg
(lower panels in Fig. 6), the mass of cool gas with
temperatures below 100, 200, and 1000~K grows with
time. Since the H$_2$ molecules are rapidly destroyed,
the gas can be cooled only by energy losses in lines of
heavy elements. Consequently, metals from the ejecta
are mixed with gas in the SN envelope, leading to a
drop in the temperature and the subsequent development
of instability. This leads to the disruption and
collapse of the envelope, and to more efficient mixing
of heavy elements. As the explosion energy grows
(second row of panels from the top of Fig. 6), the mass
of gas with temperatures below 1000~K grows rapidly
and saturates, but the gas is inefficiently cooled to
lower temperatures, and the mass of cooler gas remains
low. Note that the mass of gas with a relative
number density of H$_2$ above 5.10.4 is roughly equal
to the mass of gas with $T < 1000$~K; consequently,
this suggests that the gas is cooled primarily by the
H$_2$. Only after disruption of the envelope due to instability
and the onset of the collapse of the cavity does
the mass of gas with temperatures below 200~K, or
even below 100~K, begin to increase. This is obviously
related to mixing of the heavy elements, and therefore
to more efficient cooling in lines of heavy elements.
The case of a more massive SN with $M = 40~M_\odot$
(third row of panels) resembles the previous case—
the model with $M = 25~M_\odot$ and $E = 10^{52}$~erg. When
the initial mass of the star becomes $M = 140~M_\odot$,
the formation of molecules is efficient not only in the
strongly disrupted envelope, but also in the “lobes”
(see above and Fig. 2), so that the mass of gas in regions
with temperatures below 100, 200, and 1000~K
grows. The behavior is similar for a SN explosionwith
$M = 200~M_\odot$. Thus, cooling of the gas below 100~K in
a collapsing SN is brought about primarily by losses
in lines of metals, which are mixed with the primordial
gas during the disruption and subsequent collapse of
the SN envelope.

\section{Statistics of the distribution of gas and heavy elements}

\noindent

The heavy elements are initially concentrated in the 
gas ejected by the SN, whose metallicity is determined
by the mass of metals produced in the star (Table). It 
is obvious that this metallicity will be fairly
high: $\simgt 10~Z_\odot$. After the SN explosion, the 
expanding envelope sweeps up the surrounding primordial 
gas (this gas had a metallicity $10^{-6}~Z_\odot$ in our 
computations), and metals remain in the hot cavity. The SN
envelope is disrupted by hydrodynamical instability, and 
the metals mix with the primordial gas. However,
we can speak of mixing only when considering averaging 
over large scales. There are strictly
localized “islands” of gas enriched in metals (Fig. 2), 
which correspond to the characteristic properties of
systems with regions of intermittency -- sharp spatial 
alternation of regions with high and low metallicity,
with these values sometimes differing by an order of 
magnitude.

\begin{figure}
\center 
\includegraphics[width=12.5cm]{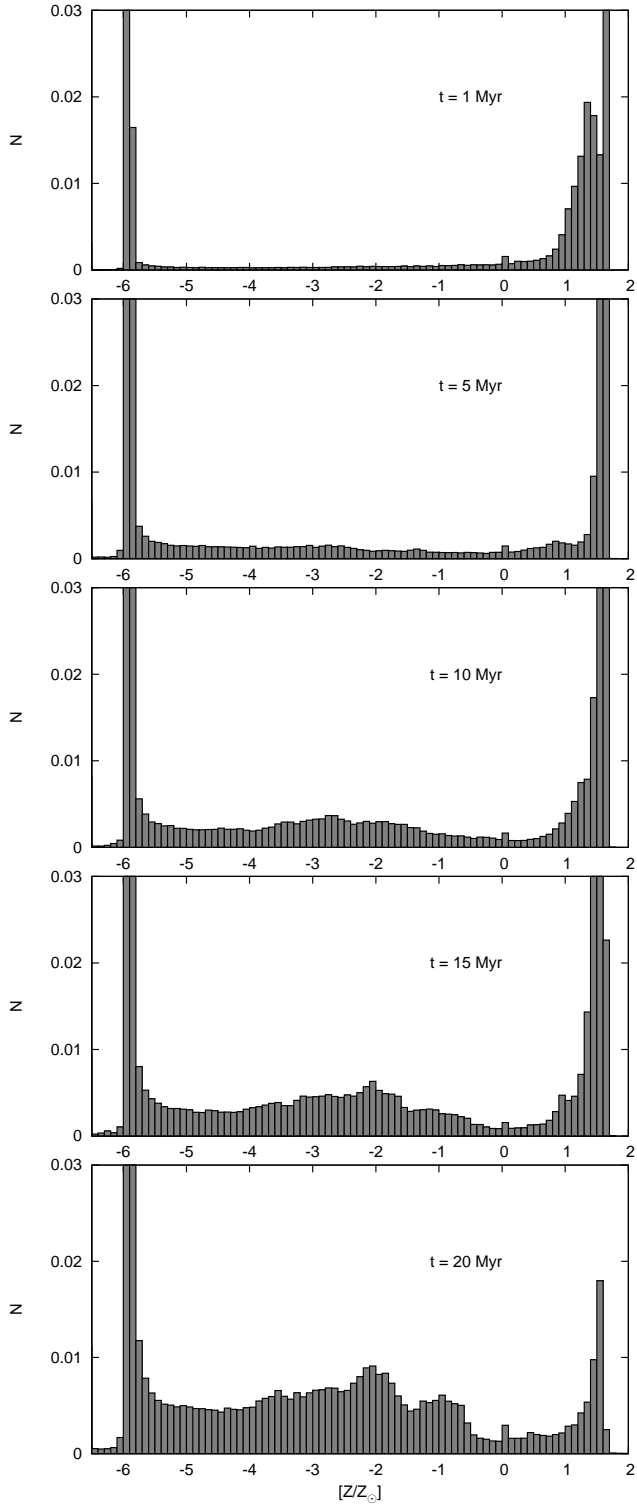}
\caption{
Distribution of the number of computation cells over gas metallicity $N(Z)$, normalized to the total number of
computation cells, for a SN explosion with energy $10^{52}$~erg and initial stellar mass 140~$\msun$ at times 
1, 5, 10, 15, and 20~Myr (from top to bottom).
}
\label{hz140e52}
\end{figure} 

\begin{figure}
\center 
\includegraphics[width=12.5cm]{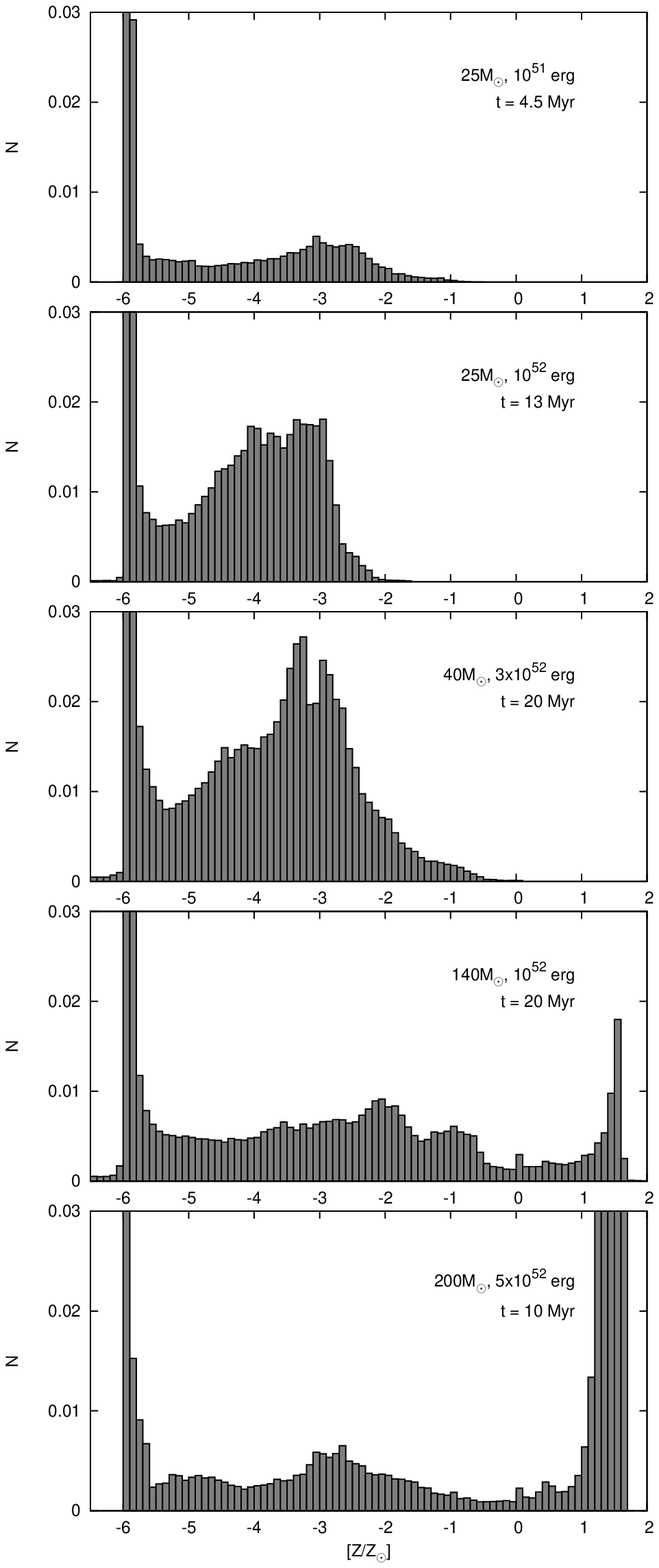}
\caption{
Distribution of the number of computation cells over the gas metallicity $N(Z)$, normalized to the total number of
computation cells, for a SN explosion with energy $E$ and initial mass $M$.
}
\label{hz}
\end{figure} 

Figure 7 shows the distribution of the number of
computation cells with various gas metallicities $N(Z)$
normalized to the total number of computation cells
for a SN explosion with energy $10^{52}$~erg and initial
mass 140~$\msun$ at times 1, 5, 10, 15, and 20 Myr.
Peaks corresponding to gas in the cavity (high metallicity)
and the unperturbed gas ahead of the shock
front ($Z = 10^{-6}~Z_\odot$) are clearly visible. With time,
the number of cells with intermediate metallicities
grows, but gas with themaximummetallicity remains
up to the final time, implying incomplete mixing of
the cavity gas. The redistribution of heavy elements
proceeds non-uniformly. A dip in the distribution at
$Z \sim Z_\odot$ and a growth in the number of cells with
metallicity $\sim 10^{-2}~Z_\odot$ can be seen. The number of
cells with low metallicities $(10^{-5} - 10^{-3})~Z_\odot$ also increases.
The gas with such metallicities is primarily
concentrated in fragments of the envelope that have
reached the periphery of the protogalaxy, $r\simgt 0.5r_{v}$.
The temperature of this gas is below 1000~K, and its
number density is above 1~cm$^{-3}$. These fragments
also contain a large amount ofmolecular hydrogen, so
that they are able to cool even further. Their properties
will be considered in more detail below.

The gas in the collapsing SN envelopes mixes
with the high-metallicity gas of the cavity due to
the action of numerous shocks and circum-sonic
waves that form during the motion of converging
flows of gas in the central region of the protogalaxy
(see the preceeding section). Therefore, the region of
gas with initially high metallicity disappears, but the
metallicity of the gas becomes appreciably higher as
the distribution of heavy elements in the protogalaxy
becomes more uniform (perfectly uniform in the case
of total mixing): $Z \sim M_{met}/M_{gas}$, where $M_{met}$ and
$M_{gas}$ are the masses of metals ejected by the SN
(Table) and the mass of gas in the protogalaxy. This is
clearly visible in Fig. 8, which presents the distribution
of the number of computation cells over the gas
metallicity $N(Z)$ normalized to the total number of
computation cells for a SN explosion with energy $E$
and initial mass $M$. In fully collapsed SN envelopes
(top three histograms), no gas with the metallicity of
the ejected gas remains at the end of the computations.
The largest numbers are displayed by cells with
metallicities in the interval $(10^{-4} - 10^{-2})~Z_\odot$ or with
the background metallicity $10^{-6}~Z_\odot$. The partially
collapsed SN envelope with $M = 140~M_\odot$ shows the
presence of the high-metallicity gas of the cavity (see
the fourth histogram from the top in Figs. 8 and 7). In
an expanding SN remnant with $M = 200~M_\odot$, we can
distinguish the background gas with $Z=10^{-6}~Z_\odot$,
hot cavity gas in which $Z \simgt 10~Z_\odot$, and gas in enriched
envelope fragments and thread-like structures
(Fig. 3) with metallicities $Z \sim 10^{-2}~Z_\odot$.

Although the distribution of the number of cells
over metallicity correctly reproduces the variations in
the character of the spatial distribution of metals with
time, it gives an incorrect impression of the filling
factors in the regions with various metallicities, and
consequently of the masses of metals contained in
these regions. This is associated with the cylindrical
geometry of our model: the cells located a distance
$r$ from the cylinder axis correspond to proportionally
larger volumes. An example of the dependence of the
mass and volume of the gas with metallicities above
a specified level and in metallicity intervals after a
SN explosion with initial mass 140~$\msun$ and energy
$10^{52}$~erg was shown in Fig. 4, and this dependence
was described in the previous section. Here, we list
the main properties of this dependence characteristic
for collapsing SN envelopes, rather than presenting
figures for all our models. First, the mass of enriched
gas with metallicities $[Z]>-5$ grows with time, and
approaches a saturation value. Second, the gas in
the central region of the protogalaxy continues to
have metallicities $[Z]>-2$, and the gas in fragments
is not enriched above $[Z]\sim-2$. Third, the volume
occupied by high-metallicity gas ($[Z]>0$) at the onset
of the evolution of the SN envelope grows, then
decreases, corresponding to the initial expansion and
subsequent compression of the hot cavity. The mass
of such gas remains nearly constant with time, and
falls off after the onset of mixing of the cavity gas and
the incident gas in the collapsing envelope. Finally,
the largest amount of gas by mass and volume is that
with intermediate metallicities $-3.5<[Z]<-2$.

Figure 9 presents the "metallicity-density" diagram for all our models. 
A typical feature of the distibutions is a incomplete mixing with an 
exceptionally high spread of 
metallicities in the whole region covered by the remnant: $-5<{\rm [Z]}<+1$. 
It is worth stressing that incomplete mixing 
takes place not only in first protogalaxies, but in the Galactic interstellar 
and intergalactiic media as well [43-45]. 
The isolines show the 
mass of gas with the indicated density and metallicity. 
It is readily seen that independent on model (e.g. pre-supernova 
density profiles, SN energy, final time etc.) the peak mass
($M\sim100-1000\msun$) is located at metallicities $-3.5<[Z]<-2$. 
This circumstance allows us to conclude that incomplete mixing of 
metals in first SN remnants provide metallicities $-3.5<{\rm [Z]}<-2$ 
typical for the old stellar population of the Milky Way halo. 
However, because the gas mass in the range  $[Z]<-3.5$ drops very fast 
this incomplete mixing process cannot explain metallicities of  extremely 
metal-poor stars which are typically below $[Z]<-3.5$ [46-51] (see in particular [52]).  
This requiers a separate study. 

\begin{figure}
\center 
\includegraphics[width=16cm]{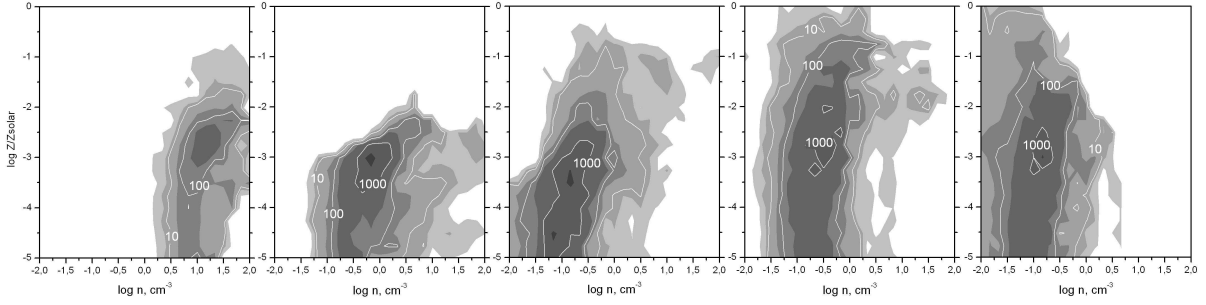}
\caption{
The "metallicity-density" diagram in gas at final time after SN explosion with 
the energy $E$ and 
the progenitor  mass $M$ (left to right):
$t = 4.4$~Myr for a star with $M=25~M_\odot$ and a SN energy $E=10^{51}$~erg,
$t = 13$~Myr for $M=25~M_\odot$ and $E=10^{52}$~erg,
$t = 20$~Myr for $E=40~M_\odot$, $E=3\times 10^{52}$~erg,
$t = 20$~Myr for $M=140~M_\odot$, $E=10^{52}$~erg,
$t = 16$~Myr for $M=200~M_\odot$ and $E=5\times 10^{52}$~erg.
The isolines show those cells with given density and metallicity where the 
mass content equals the indicated value.
}
\label{rhoz}
\end{figure} 

Figure 3 shows that the SN envelope is strongly
disrupted, and its fragments contain cool, denise gas
enriched in metals. These fragments are $\sim 1-10$~pc
in size, and have densities a factor of ten or more
higher and temperatures a similar factor lower than
the surrounding regions. The further from the center
of the protogalaxy, the lower the density of the gas in
the fragments. Thus, the fragments remaining after
the disruption of the envelope of a SN with initial
mass 140~$\msun$ and energy $10^{52}$~erg turn out to be less
dense than in the case of a SN with $M = 25~M_\odot$ and 
$E = 10^{51}$~erg. This can easily be explained by
the value of the surrounding gas density at the onset
of the disruption of the SN envelope: the lower the
explosion energy, the smaller the size of the envelope
when it begins to disrupt. The fragments possess
primarily supersonic velocities. Gas densifications
associated with the propagation of the shock front
can be noted around the fragments at the periphery
of the protogalaxy. The fragments moving toward the
center also have supersonic velocities of the order of
several km/s. The characteristic time for the disruption
of such fragments when they are ripped apart
by the Kelvin-Helmholtz instability is several million
years. On the other hand, the fairly high metallicity,
$[Z]\sim -3$, and H$_2$ number density $\sim 2\times 10^{-3}$, of the
fragments facilitate rapid cooling of the gas over a
time scale shorter than those for their disruption or
destruction. However, a substantial number of fragments
in the central region are disrupted by collisions,
hindering efficient cooling and the formation of protostellar
condensations.

The identification of distinct fragments with densities
higher than a specified level (using the method
described in [52, 60]) showed that the masses of these
fragments are smaller than the Jeans mass. A central,
cool, denise fragment gradually forms in collapsing
SN envelopes, which contain a larger amount of
heavy elements than similar fragments at the protogalaxy
periphery. The mass of this fragment grows
due to the infalling gas, which rapidly cools due to
energy losses in lines of heavy elements. We ceased
our computations when a gravitational-compression
regime was established in the central region of the
protogalaxy, i.e., when the gas density reached $n\sim 10^6$~cm$^{-3}$
(the reasons for this are indicated above).
By this time, the mass of the central fragment with a
density above $10^{-3}$~cm$^{-3}$ reaches $\sim 70~\msun\sim 1/4M_J$
for a SN with $M = 25~M_\odot$ and $E = 10^{51}$~erg, and
will probably continue to grow further. The mass
of this fragment is $\sim 2\times 10^3~\msun \simgt 100 M_J$ for a SN
with $M = 25~M_\odot$ and $E = 10^{51}$~erg, as well as a
SN with $M = 40_\odot$ and $E = 3\times 10^{52}$~erg. The
mean density, temperature, and metallicity of the
gas in the fragment are approximately $\sim 10^5$~cm$^{-3}$, 
$\sim 40$~K, and $\sim$0.01~$Z_\odot$. It is obvious that the birth
of a group of stars with various masses (including
some with masses comparable to the solar mass)
will subsequently be possible in this region of gas,
whose lifetimes will be several billions of years. These
could live to the current epoch and be observed in the
Galaxy. Moreover, the birth of stars with high masses
$\simgt 10\msun$ is probable, whose ultraviolet radiation and
explosions as SN could faciliate subsequent mixing
of heavy elements. Finally, in the case $M = 140~M_\odot$
and $E = 10^{52}$~erg, the mass of the central fragment
turns out to be substantially lower than the corresponding
Jeans mass, although the number density at
the center reaches $n\sim 10^6$~cm$^{-3}$. It is possible that
such a central fragment could subsequently become
gravitationally bound.

\section{Discussion}

\noindent

Some consequences of SN explosions (such as
the formation of low-mass galaxies with virtually no
gas) and possible problems with our computations (in
particular, failure to take into account CO and OH
molecules) are discussed above. All the computations
for which results are presented above assumed that
a single massive star was born in the initial phase
of the protogalaxy, due to inefficiency of fragmentation
in that stage, as was manifest in the numerical
computations of [6]. However, the conditions for the
formation of a large number of protostellar fragments
could well arise in rotating protogalaxies or strongly
turbulent gas [7--10]. Consequently, numerous SN
explosions are, in principle, possible in the first protogalaxies.
Of course, the initial surrounding gas
would be ionized by radiation from such a cluster of
first-generation stars, whose masses would likely be
fairly high due to the low opacity of the primordial
gas [3]. Therefore, the gas in, and probably beyond,
the protogalaxy will be ionized, although this depends
on the mass function of these first stars. Here, we
should bear in mind that such a cluster will not contain
many first stars (the mass of cool gas in a low-mass
protogalaxy cannot be high), so that the masses
of stars in the cluster will be distributed randomly
over a wide interval, e.g., $M\sim 10-1000~\msun$. It seems
unlikely that most of these will be stars with masses
of 140--260~$\msun$, capable of exploding as high-energy
supernovae. Since some of these stars could form
black holes (40--140~$\msun$), they could participate only
in the formation of the ionization zone. The few
stars that explode as supernovae will probably explode
nearly simultaneously, forming an envelope equivalent
to that for a single supernova with the same total
power.

Let us suppose that the stars in such a cluster
were born approximately simultaneously. In this case,
the envelope of a SN explosion of a star with mass
140~$\msun$ and energy $10^{52}$~erg will still be expanding
$\sim$5-6~Myr after the explosion (Fig. 1), so that a
star with mass 25~$\msun$ (with a lifetime of 6.5~Myr)
could explode before the envelope of the preceding SN
cools and collapses. In this case, the shocks from
the second and preceding supernovae will propagate
through highly enriched gas (or capture this gas, if
the parent stars are appreciably separated in space),
stimulating mixing of the gas during the collision at
the inner side of the first envelope.

If the birth of stars in the cluster takes place
over an appreciable time, the envelope from the first
supernova could have time to potentially cool and
fragment. The influence of the ionization fronts from
other stars and the subsequent action of the shock
front on other supernovae could then both disrupt
fragments in the first envelope and stimulate their
compression, i.e., stimulate star formation. The efficiency
of mixing of metals will obviously increase in
this case due to the additional energy supplied (see
the Introduction). Taking into account our comments
about mixing and the results presented in the previous
sections, we conclude that, after the explosions of
the first supernovae, it is probably difficult to detect
a clear relationship between the metallicities of stars
and the generation in which they were born. Thus,
both stars with metallicities close to the solar value
(incomplete mixing) and metal-poor stars (complete
mixing) could be born in gas enriched by the first
stars.

Recently, the evolution of the first stars with rotation
has been considered [82]. The winds from such
stars could be fairly powerful, and could contain a
significant amount of metals in later stages of the
star’s lifetime [82]. Thus, rotating stars could exert
a significant chemical effect on the surrounding gas,
even before their explosion as supernovae. The efficiency
of this effect depends on the mass of metals in
the stellar wind and the velocities of gas outflows from
the stellar surfaces [83].

Finally, dust particles can form in SN remnants
[84–87], which can play a appreciable role
in star formation. First, dust particles substantially
increase the opacity of a medium, facilitating efficient
fragmentation and a transition to the birth of low-mass
stars [88], compared to the case of gas without
dust [34-37]. Second, the formation of molecular
hydrogen is appreciably enhanced in the presence of
dust [89], likewise leading to more efficient cooling of
the gas and facilitating the birth of lower-mass stars.
However, the transport and mixing of dust could differ
substantially from the transport of gas particles, since
dust is appreciably heavier and has appreciably lower
charge-to-mass ratios than ions; in addition, dust will
be disrupted behind shock fronts [90], and can grow
in cool, dense clouds [91]. Therefore, the questions
of the transport of dust, and also of its formation in
SN, remain open [84, 87, 92], and requires a separate
discussion.

\section{Conclusions}

\noindent

We have considered the dynamical, thermal, and chemical evolution of gas in supernova envelopes
from first stars with masses$M_*\sim 25-200~M_\odot$ in protogalaxies with mass $M\sim 10^7~M_\odot$ 
at redshifts $z = 12$, and analyzed the efficiency of mixing of heavy elements produced in these 
first stars. The SN explosions occur inside the ionization zone formed by the star. In particular, 
we have shown the following.

\begin{enumerate}
 \item  During SN explosions with high energies ($E\simgt 5\times 10^{52}$~erg), an appreciable amount 
 of gas can be ejected from the protogalaxy, but nearly all the metals produced remain inside the galaxy. 
 The hot, enriched gas occupies a central volume with a radius up to half the virial radius for a protogalaxy 
 with $M\sim 10^7~M_\odot$; the cooling time in this hot gas is of order several million years -- comparable to 
 or longer than the time for the loss of a substantial fraction of the gas mass (more than 50\%). Thus, 
 protogalaxies in which high energy SN explosions occur may be transformed into "dark" objects with virtually 
 no gas.
 \item During SN explosions with lower energies ($E\simlt 3\times 10^{52}$~erg), essentially no gas and heavy 
 elements are lost from a protogalaxy with $M\sim 10^7~M_\odot$. During the first 1-3~Myr, gas and heavy 
 elements are actively carried from the central region of the protogalaxy: up to 90\% of the mass of metals 
 is carried beyond a radius of 0.05 times the virial radius during a SN explosion with an initial mass of
 25~$\msun$ and energy $E = 10^{51}$~erg, and beyond a radius of 0.1 times the virial radius during a SN 
 explosion with an initial mass of 40~$\msun$ and energy $E = 3\times 10^{52}$~erg or a SN explosion with 
 an initial mass of 140~$\msun$ and energy $E = 10^{52}$~erg. However, during the subsequent evolution,
 an appreciable fraction of the mass of metals returns to the center as the hot cavity is cooled and
 the envelope collapses.
 \item High-energy supernovae ($E\simgt 5\times 10^{52}$~erg) are characterized by a low efficiency of mixing 
 of metals. Heavy elements are located in a small volume in the disrupted envelope (compared to the entire volume 
 of the envelope), and the bulk of the metals remain inside the hot, rarified cavity.
 \item  The efficiency of mixing of heavy elements for lower-energy supernovae ($E\simlt 3\times 10^{52}$~erg) 
 is appreciably higher than for SN with energies $E\simgt 5\times 10^{52}$~erg due to the disruption of the hot
 cavity during the collapse of the SN envelope. However, a clear distinction between enriched and unenriched
 (primordial) regions is observed; i.e., the mixing remains incomplete. 
 \item During the collapse of the SN envlope, the hot, metal-rich gas of the cavity mixes with the cool, primordial
 (metal-poor) gas of the envelope. Therefore, the metallicity is appreciably higher in the central region of the
 protogalaxy ($[Z]\sim -1$ to $0$) than in peripheral regions ($[Z]\sim -2$ to $-4$). The bulk of the enriched gas
 has metallicities $[Z]\sim -3.5$ to $-2.5$. 
 \item The enriched regions in the SN envelope are mainly characterized by fairly high metallicities $[Z]\simgt -3$
 (higher than the critical value $[Z]_{cr}\sim -3.5$) and high H$_2$ abundances ($x({\rm H_2})\simgt 10^{-3}$). The
 densities of such regions turn out to be substantially higher, and their temperatures substantially lower, than the
 mean values for the surrounding gas; however, their masses are appreciably lower than the Jeans mass, except in the
 central region of the protogalaxy. The ambient enriched gas accretes efficiently into the central region, and the
 birth of stars with metallicities close to those characteristic of modern stars in the Galaxy is very probable in
 this region.
\end{enumerate}

\section{Acknowledgements}

\noindent

This work was supported by the Russian Foundation
for Basic Research (projects 09-02-00933,
11-02-00621, 11-02-01332, 11-02-90701, 11-02-97124, 12-02-90800), 
the “Dinasty” Non-commercial Foundation,
the Austrian Science Foundation (FWF; grant
M 1255-N16), and the Ministry of Education and
Science of the Russian Federation (state contract P-
685). The computations were carried out on clusters
of the Computational Center of the Southern Federal
University.



\end{document}